\documentclass[11pt]{article}

\usepackage{amsmath, amssymb, amsfonts, amsthm}
\usepackage{algorithm}
\usepackage[noEnd=true]{algpseudocodex}
\usepackage{xcolor}
\usepackage{hyperref}
\usepackage{qcircuit}
\usepackage{subcaption} 
\usepackage{biblatex}
\usepackage{braket}
\usepackage{enumitem}
\newtheorem{theorem}{Theorem}[section]

\newtheorem{proposition}[theorem]{Proposition}

\def\qed{\hfill $\Box$\medskip}

\def\ra{{\rangle}}

\def\ZZ{{\mathbb{Z}}}
\def\IZ{{\ZZ}}

\def\cF{{\cal F}}
\def\cG{{\cal G}}

\begin{document}
\openup .8\jot

\title{Minimum-Spanning-Tree Tomography of Sparse Quantum States With and Without Entanglement}
\author{Chi-Kwong Li, Kevin Yipu Wu, and Zherui Zhang\\
Department of Mathematics, College of William \& Mary, Williamsburg, VA 23187, USA.\\
ckli@math.wm.edu, ypwk@uw.edu, zzhang30@wm.edu}

\maketitle
\begin{abstract}
Many quantum states arising in algorithms and physical systems occupy only a small, structured subset of the exponentially large Hilbert space, yet standard quantum state tomography fails to exploit this structure.
We present an efficient circuit-based tomography framework for pure quantum states that are sparse in a computational basis.
For an $n$-qubit state supported on $k$ basis elements, the protocol reconstructs all amplitudes using $1 + 2(k-1)$ measurement settings.
The method admits both entanglement-assisted and entanglement-free implementations, enabling explicit tradeoffs between two-qubit gate usage and statistical noise.
We derive bounds on the required number of CNOT gates from the combinatorial structure of the state support and analyze their effect on reconstruction infidelity.
The framework extends naturally to closed-system process tomography and is validated via numerical simulations using Qiskit.
\end{abstract}


\section{Introduction}

Quantum tomography is a technique for reconstructing the complete quantum state of a physical system by performing measurements on multiple identically prepared copies of the state. Specifically, by measuring a quantum system in various bases, one gathers statistical data from which the state can be inferred. Quantum state tomography (QST) thus serves as an essential verification tool for experimental quantum information science, enabling researchers to confirm that quantum devices perform as intended and providing insight into quantum state properties such as coherence, entanglement, and purity. For comprehensive background discussions and additional context, readers are referred to \cite{blumekohoutOptimalReliableEstimation2010,flammiaQuantumTomographyCompressed2012,hradilQuantumstateEstimation1997,jamesMeasurementQubits2001,nielsenQuantumComputationQuantum2010}.

We study the theory and implementation of quantum state tomography for pure states, which are important for the following reasons.
    They provide the most direct description of coherent quantum evolution, as quantum algorithms are implemented by applying unitary operations to an initial pure state and extracting information through measurement at the end of the computation.
    Consequently, understanding and verifying pure states is essential for characterizing the behavior of quantum devices and algorithms. Moreover, pure states are often simpler to prepare and control experimentally than mixed states, and many experimental platforms naturally operate close to the pure-state regime, particularly over short coherence times.
For these reasons, quantum tomography of pure states is an important and widely studied topic in quantum physics and quantum information science; see
\cite{altepeterPhotonicStateTomography2005, blumekohoutOptimalReliableEstimation2010, childsQuantumInformationPrecision2000, flammiaQuantumTomographyCompressed2012} and their references.

In \cite{blumekohoutOptimalReliableEstimation2010, emersonScalableNoiseEstimation2005, flammiaQuantumTomographyCompressed2012}, the authors discuss various aspects of the complexity of pure state tomography using the circuit model, including the scalability of the methods, the efficiency of the algorithms used for state reconstruction, and the role of entanglement in determining the usefulness of quantum states as computational resources. 
The complexity can range from exponential to polynomial in $n$, depending on the specifics of the problem. In particular, the complexity of determining an $n$-qubit state using the circuit model for pure state tomography depends on the number of gates applied to the state to obtain measurement data. 

We are motivated by the prevalence of sparse quantum states in both theoretical models and quantum algorithms. While general quantum state tomography scales exponentially with system size, many states of interest in quantum information science occupy only a small, structured subset of the computational basis \cite{bellanteQuantumSparseRecovery2025, schreiberTomographyParametrizedQuantum2025, xieReconstructingQuantumStates2023}. In such cases, the physically relevant information is concentrated in a limited number of nonzero amplitudes, rendering full tomographic reconstruction both inefficient and conceptually misaligned with the s
tructure of the problem. In optimization-oriented quantum algorithms, such as a adiabatic or variational schemes, the underlying Hilbert space grows exponentially with the problem size, yet the states produced during computation, especially low-energy or near-optimal states, are frequently supported on a small subset of basis configurations \cite{kieuTravellingSalesmanProblem2019, padmasolaSolvingTravelingSalesman2025, salesAdiabaticQuantumComputing2023}. These states may correspond to a single optimal solution or a small superposition of candidate solutions, rather than a generic high-entropy state. In such regimes, the goal of tomography is to efficiently identify and characterize the dominant amplitudes that encode the computational outcome. This perspective motivates tomographic protocols whose resource requirements scale with the intrinsic structure of the state rather than with the full Hilbert space dimension \cite{bellanteQuantumSparseRecovery2025, schreiberTomographyParametrizedQuantum2025}. By exploiting sparsity as a structural property of the target state, one can design reconstruction methods that are better matched to both the physics of the system and the computational tasks of interest. 

In this work, we introduce an efficient quantum state tomography scheme that leverages the sparsity structure of quantum states in the computational basis.
Section~\ref{sec:notation_and_preliminaries} establishes the notation and background needed for the development of the method.
Section~\ref{sec:algorithm} presents the tomography protocol in detail, together with several noise-mitigation techniques.
In Section~\ref{sec:structural_results_and_cnot_guarantees}, we provide a structural theorem that bounds the number of entangling or mixing operations required by the scheme.
Section~\ref{sec:regime_selection} offers guidance on selecting an operational mode suitable for different experimental regimes.
Section~\ref{sec:qprocess_tomo_closed_system} extends the framework to closed-system process tomography.
Finally, Section~\ref{sec:numerical_simulations_and_results} reports the results of numerical simulations demonstrating the performance of the method.

\section{Notation and Preliminaries}
\label{sec:notation_and_preliminaries}
An $n$-qubit pure quantum state $|\psi\rangle$ can be written as a unit vector in the Hilbert space $\mathcal{H} = (\mathbb{C}^2)^{\otimes n}.$ 
The standard basis for $\mathcal{H}$ is given by the set $\{|k\rangle : k \in \{0,1\}^n\}$, where each $|k\rangle$ represents a basis state corresponding to a binary string of length $n$. 
Any expansion of the state $|\psi\rangle = \sum_{k \in \{0,1\}^n} a_k \, |k\rangle$ expresses $|\psi\ra$ in terms of its probability amplitudes $a_k \in \mathbb{C}$, which satisfy $\sum_{k} |a_k|^2 = 1.$ 
In this context, an \textit{entry} of $|\psi\rangle$ refers to one of these amplitudes. 
We say that $|\psi\rangle$ is \emph{sparse} if only a subset (say, $k$ out of $2^n$) of the amplitudes are nonzero. 
Let 
\[
H  = \frac{1}{\sqrt 2}\begin{pmatrix} 1 & 1 \\ 1 & -1\end{pmatrix},
\qquad
V = \frac{1}{\sqrt 2}\begin{pmatrix} 1 & i \\ 1 & -i\end{pmatrix}
\]
denote the Hadamard gate and a single-qubit unitary operator acting on a qubit, respectively (which, for instance, can be viewed as the Hadamard gate followed by an appropriate phase shift).

\section{Algorithm}
\label{sec:algorithm}

In this section, we describe a general tomography scheme making using of the sparsity pattern of the nonzero entries of $|\psi\ra$. In subsections \ref{ssec:full_support} and \ref{ssec:one_two_sparse_states}, we describe simpler examples of states with full support or one/two-sparsity and how we may use entanglement to adjust for certain sparse patterns. In subsection \ref{ssec:partial_mixing}, we describe an entanglement-free alternative. In subsection \ref{ssec:general_sparse_tomography}, we give the full sparse tomgraphy algorithm.

\subsection{Full Support States}
\label{ssec:full_support}
The following result is known. We give a short proof for easy reference. 
\begin{proposition}
Suppose $|\psi\ra$ is $n$-qubit state with no zero entries. 
Then $|\psi\ra$ can be reconstructed using the measurements of $|\psi\ra, H_{n-1}|\psi\ra,  \cdots H_0|\psi\ra, V_{n-1}|\psi\ra, \dots, V_0|\psi\ra$.
\end{proposition}

\begin{proof}
    We prove the result by induction on $n$.
    Since $e^{i\phi} |\psi\ra$ and $|\psi\ra$ represent the same quantum state, we may assume that $(x_0,x_1)  = (\cos\theta, \sin \theta e^{i\phi})$ with $\theta \in (0, \pi/2)$ and $\phi \in [0, 2\pi)$. 
    The measurement of $|\psi\ra$ will determine $|x_0|^2, |x_1|^2$.
    The measurements of $H|\psi\ra$ and $V |\psi\ra$ will determine $(|x_0+x_1|^2, |x_0-x_1|^2)$ and $(|x_0+ix_1|^2, |x_0-ix_1|^2)$.
    So, $x_1$ can be determined by $|x_0 + x_1|^2 = 1 + 2 \Re x_0\bar x_1$ and $|x_0+ix_1|^2 = 1 - 2\Re x_0 (i\bar x_1)$.
    
    Suppose $n > 1$, and the result is proved for $(n-1)$-qubit states.
    Let $|\psi\ra$ be an $n$-qubit state and assume $|\psi\ra = \cos\theta |0\ra |\psi_1\ra + \sin\theta e^{i\phi} |1\ra |\psi_2\ra$ with $\theta \in (0, \pi/2)$, $\phi\in [0, 2\pi)$, so that the first entry of $|\psi_j\ra$ is positive for $j = 1,2$.
    Then the measurements of $H_\ell |\psi\ra, V_\ell |\psi\ra$ for $\ell = n-2, \dots, 0$, will determine $|\psi_1\ra, |\psi_2\ra$ by the induction assumption.
    Finally, the measurements of $H_{n-1}|\psi\ra$ and $V_{n-1}|\psi\ra$ will determine $\cos\theta$ and $e^{i\phi}$.
\end{proof}

\subsection{Two-term superpositions}
\label{ssec:one_two_sparse_states}
Consider $|\psi\ra$ that is 2-sparse, a state with two nonzero entries. These nonzero entries may form different sparse patterns, which can make the tomography process easier or harder. 

Two-term superpositions occupy a particularly important role in quantum information science. 
They represent the simplest class of genuinely quantum states that exhibit interference, relative phase, and entanglement, while remaining analytically and experimentally tractable. 
Many foundational phenomena, such as Bell nonlocality, quantum teleportation, and elementary entangling gates, are naturally expressed in terms of states supported on exactly two computational basis states.

We first measure $|\psi\ra$ in the computational basis. This measurement gives us estimates of the amplitudes, up to quantum shot noise. We can decide which entries are zero or nonzero by comparing their measured probabilities with a small threshold value. If we find that there is only one nonzero amplitude, then we can assume that $|\psi\ra = |q_{n-1} \cdots q_0\ra$ with $q_j \in \{0,1\}$ for $j = 0, \dots, n-1$. Throughout the rest of this paper, we assume that the computational-basis support can be reliably identified from sampling statistics, and we focus on the subsequent phase- and amplitude-reconstruction problem conditioned on this support.

Suppose we find two nonzero amplitudes in positions $i$ and $j$. We write these locations as binary strings $i = (i_{n-1}, \dots, i_0)$ and $j = (j_{n-1}, \dots, j_0)$ in $\mathbb{Z}_2^n$. The \textit{Hamming distance} $h(i,j)$ counts the number of digits that are different between the two strings. This is equal to the number of nonzero entries in the vector $(i_{n-1} \cdots i_0) + (j_{n-1} \cdots j_0)$ where addition is in $\mathbb{Z}_2$. We have two cases, \(h(i,j) = 1\) and \(h(i,j) > 1h\):
\begin{itemize}
    \item If $h(i,j) = 1$ so that $i_p \ne j_p$ and $i_\ell = j_\ell$ for all $\ell \ne p$, then we can determine $|\psi\ra$ using the measurements $H_p |\psi\ra$ and $V_p |\psi\ra$ if the amplitude at position $i$ is positive and real by fixing the overall global phase. 
    \item If $h(i,j) = n> 1$, then measurements of $H_j|\psi\ra$ and $V_j|\psi\ra$ for $j = 0, \dots, n-1$ are not enough to determine $|\psi\ra$. For convenience, we will write $(i_{n-1}, \dots, i_0)$ as $(i_{n-1} \cdots i_0)$. For example, if $n = 2$, we may have $i = (00)$ and $j = (11)$. We can apply the CNOT gate defined by
    \[
        C|q_1 q_0\ra = |q_1\ra\,|q_0 \oplus q_1\ra
    \]
    After applying $C$, the state $C|\psi\ra$ has its nonzero entries at $(00)$ and $(10)$. We can then use the measurements of $H_1 C|\psi\ra$ and $V_1 C|\psi\ra$ to determine the values of the nonzero amplitudes. As before, we assume that the first nonzero amplitude of $|\psi\ra$ is real and positive by fixing the overall global phase. 
    
    If $h(i,j) = m$ we can apply $m-1$ CNOT gates to obtain a new state $|\hat\psi\ra$ whose two nonzero entries lie in positions $i$ and $\hat j$ where $h(i,\hat j) = 1$. We can then determine the amplitudes using single qubit measurements.
\end{itemize}

\noindent
In general, we have the following result.

\begin{proposition}  Suppose $|\psi\ra$ is an $n$-qubit
state with two nonzero entries at the positions labeled by two binary
sequence $i = (i_{n-1} \cdots i_0)$ and $j=(j_{n-1} \cdots j_0)$
such that $h(i,j) = m$. Then 
$|\psi\ra$ can be determined by the measurements of 
$H_\ell R |\psi\ra$ and $V_\ell R|\psi\ra$, where 
$\ell \in \{0, \dots, n-1\}$ and $R$ is a product of $m-1$ CNOT gates.
    \end{proposition}

\subsection{Partial mixing tomography}
\label{ssec:partial_mixing}
Motivated by the high failure probability associated with entangling operations,
we introduce a partial mixing approach that allows us to carry out a similar
interference-based scheme using only single-qubit gates.

Suppose the nonzero entries of $|\psi\rangle$ occupy basis states labeled by
binary strings $i = (i_{n-1}\cdots i_0)$ and $j = (j_{n-1}\cdots j_0)$.
Let $D(i,j) \subseteq \{0,\dots,n-1\}$ be the set of qubit indices where
$i$ and $j$ differ, so that $|D(i,j)| = h(i,j)$.
We choose one index $p \in D(i,j)$ as a pivot qubit, and denote
$S = D(i,j) \setminus \{p\}$, so $|S| = h(i,j)-1$.

We then apply Hadamard gates on the qubits in $S$ and either a Hadamard or
phase--Hadamard gate on the pivot:
\[
U_{i,j}^{(H)} = H_p \bigotimes_{\ell \in S} H_\ell,
\qquad
U_{i,j}^{(V)} = V_p \bigotimes_{\ell \in S} H_\ell.
\]
We call these operations \emph{partial mixing unitaries}.  They coherently spread
the amplitudes $x_i$ and $x_j$ over a $2^{h(i,j)-1}$-dimensional subspace
in such a way that, for each fixed pattern of the qubits in $S$, the pivot
qubit sees a two-dimensional superposition of the form
\[
\alpha_s |0_p\rangle + \beta_s |1_p\rangle
\propto x_i |0_p\rangle + \sigma(s) x_j |1_p\rangle,
\]
where $\sigma(s) = \pm 1$ is a known pattern-dependent sign.

By measuring $U_{i,j}^{(H)}|\psi\rangle$ and $U_{i,j}^{(V)}|\psi\rangle$ in
the computational basis and grouping outcomes according to the value of
$\sigma(s)$, one obtains quantities proportional to
\[
|x_i + x_j|^2,\quad |x_i - x_j|^2,
\qquad
|x_i + i x_j|^2,\quad |x_i - i x_j|^2.
\]
From these four real numbers, the magnitude ratio and relative phase between
$x_i$ and $x_j$ can be recovered, once one amplitude is fixed to be real and
positive by a global phase convention.  No entangling gate is required
whenever $h(i,j)\ge 1$; the number of single-qubit gates scales linearly with
the Hamming distance $h(i,j)$ and is independent of the total number of
qubits $n$.

More generally, partial mixing allows us to handle pairs with moderately
large Hamming distance without first reducing their distance via CNOT
operations.  However, spreading the state's probability mass across
$2^{h(i,j)-1}$ basis states increases the sensitivity of the protocol to
quantum projection noise and readout errors, and there is a tradeoff between
the size of the mixed subspace and the reduction in two-qubit gate errors.
See Section~\ref{sec:regime_selection} for a quantitative analysis of the
tradeoff between CNOT noise and quantum shot noise.

\subsection{Sparse Tomography Algorithm using a Graph Theory}
\label{ssec:general_sparse_tomography}

Suppose an $n$-qubit state $|\psi\ra = (x_0, \dots, x_{N-1})^t$ has $k$ nonzero entries $x_{\ell_1}, \dots, x_{\ell_k}$, where $N = 2^n$. Let $\cG$ be the graph with vertices $\ell_1,\dots, \ell_k$, and  the Hamming distance $h(\ell_u,\ell_v)$ as the weight of the edge connecting the two vertices $\ell_u, \ell_v$. Then there is a subgraph of $\cG$ with $k-1$ edges connecting all the vertices such that the sum of the weights of these $k-1$ edges is minimum. Such a subgraph of $\cG$ is called a minimal spanning tree (MST) \(T\) of $\cG$. There are standard algorithms to construct a minimal spanning tree; e.g., see \cite{kruskalShortestSpanningSubtree,primShortestConnectionNetworks1957}. 

Each edge of $T$ corresponds to a two-dimensional interference problem between
a pair of amplitudes $(x_i,x_j)$. Such an edge may be resolved in one of two
ways: 

\begin{enumerate}[label=(\roman*)]
    \item by reducing the Hamming distance via CNOT gates and performing standard
single-qubit interference (the entanglement-assisted, or ENT, scheme), or
    \item by applying partial mixing on the differing qubits and performing
interference on a chosen pivot qubit without entanglement (the PM scheme).
\end{enumerate}

We refer to any unitary operation that enables the determination of an
amplitude ratio $x_i/x_j$ for a given edge $(i,j)$ of the support graph as an
\emph{edge-resolution unitary}. The MST therefore determines which amplitude ratios must be learned,
while the choice between ENT and PM determines how each ratio is
implemented at the circuit level. We have the following.

\bigskip\hrule\hrule\medskip\noindent
{\bf A tomography algorithm for states with zero entries}
\medskip
\hrule\hrule\medskip

\noindent
Let $|\psi\ra = (x_0, \dots, x_{N-1})^t$ be an $n$-qubit state, where $N = 2^n$.
Measure $|\psi\ra$ to determine the magnitudes of the nonzero entries, say, $k$ of them
at positions labeled by binary sequences $\ell_1, \dots, \ell_k \in \IZ_2^n$. Let $\cG$ be the graph with  vertices $\ell_1,\dots, \ell_k$, and 
the Hamming distance $h(\ell_u,\ell_v)$ as the weight
of the edge connecting the 
two vertices $\ell_u, \ell_v$. 
Determine a minimal tree $T$ of $\cG$, and let 
$p_j$  be the number of  edges of $T$ with weight  
$j = 1, \dots, r$, where
$r < n$ and $p_1 + \cdots + p_r = k-1$.

\medskip\noindent{\bf Step 1.} Let 
$\cF_1$ be the subgraph of $T$ formed by the vertices of $T$
and the $p_1$ edges of weight 1. 
For each edge in $\cF_1$, say, joining
vertices $i = (i_{n-1} \cdots i_0)$ and $j=(j_{n-1} \cdots j_0)$, 
which differ only in one entry, 
use two single qubit gates of the form $H_u, V_u$ with $u \in \{0, \dots, n-1\}$ 
to determine the ratio $x_i/x_j$. 
If $\cF_1$ has $s$ connected components 
and $|\psi\ra = |\psi_1\ra + \cdots +|\psi_s\ra$, where
$|\psi_1\ra, \dots, |\psi_s\ra$ 
contains the nonzero entries of 
$|\psi\ra$ with entries in the positions labeled by the
different components of $\cF_1$, then
each $|\psi_1\ra, \dots, |\psi_s\ra$ is determined
up a unit multiple. Let $\ell = 1$ and go to Step 3.

\medskip\noindent
{\bf Step 2.} If $\cF_\ell$ has only one connected component, then we are done.
Otherwise,  
let $t$ be the smallest positive integer
such that $p_{\ell+t} > 0$
and let $\cF_{\ell+t}$ be the subgraph of $T$  containing $\cF_\ell$
together with the $p_{\ell+t}$ edges in $T$ having weight $\ell+t$ of $T$. 
For each edge in $\mathcal{F}_{\ell+t}$ of weight $\ell+t$ joining vertices
$i=(i_{n-1}\cdots i_0)$ and $j=(j_{n-1}\cdots j_0)$, one determines the ratio
$x_i/x_j$ using one of the following implementations:

\begin{itemize}
\item[(ENT)] Apply a sequence $R$ of $\ell+t-1$ CNOT gates to reduce the
Hamming distance to one, followed by single-qubit measurements of the form
$H_u R|\psi\rangle$ and $V_u R|\psi\rangle$.

\item[(PM)] Choose a pivot qubit among the $\ell+t$ differing positions and
apply partial mixing unitaries consisting of Hadamard gates on the remaining
$\ell+t-1$ differing qubits and either $H$ or $V$ on the pivot qubit, as
described in Section~\ref{ssec:partial_mixing}.
\end{itemize}

If $\cF_{\ell+t}$ has $s$ connected components 
and $|\psi\ra = |\psi_1\ra + \cdots +|\psi_s\ra$
such that $|\psi_1\ra, \dots, |\psi_s\ra$ 
contains the nonzero entries of 
$|\psi\ra$ with entries in the positions labeled by the
different components of $\cF_{\ell+t}$, then
each $|\psi_1\ra, \dots, |\psi_s\ra$ is determined
up to a unit multiple. Let $\ell = \ell + t$ and repeat Step 3.

\medskip\hrule\hrule\bigskip

Following the procedures in the above algorithm, we have the following.

\begin{proposition}
Suppose an $n$-qubit state $|\psi\rangle = (x_0,\dots,x_{N-1})^t$ has $k$ nonzero
entries $x_{\ell_1},\dots,x_{\ell_k}$, where $N=2^n$.
Let $\mathcal{G}$ be the graph with vertices $\ell_1,\dots,\ell_k$ and edge
weights given by the Hamming distance $h(\ell_u,\ell_v)$.
Suppose $\mathcal{G}$ has a minimal spanning tree $T$ with $p_j$ edges of weight
$j$ for $j=1,\dots,r$, where $r<n$ and $p_1+\cdots+p_r=k-1$.

Then the nonzero entries of $|\psi\rangle$ can be determined using measurements
of $U_1|\psi\rangle,\dots,U_{2m}|\psi\rangle$ with $m\le k$, where for each
$j=1,\dots,m$, the pair $(U_{2j-1},U_{2j})$ consists of edge-resolution unitaries
that determine the ratio $x_{\ell_u}/x_{\ell_v}$ for some edge
$(\ell_u,\ell_v)\in T$.

If all edges are resolved using the entanglement-assisted (ENT) implementation,
then each pair $(U_{2j-1},U_{2j})$ may be taken to be of the form
$(H_{u_j}R_j,\,V_{u_j}R_j)$, where $R_j$ is a product of CNOT gates.
In this case, the total number of CNOT gates used across all measurements is
bounded above by
\[
2(p_2 + 2p_3 + \cdots + (r-1)p_r).
\]
\end{proposition}

By the above algorithm, an $n$-qubit state with sparse support can be
reconstructed efficiently.  The minimal spanning tree determines the set of
amplitude ratios that must be resolved, while the choice of edge-resolution
unitaries (ENT or PM) determines the circuit-level resource requirements.

If $\mathcal{F}_1$ in Step~1 has only one connected component, then all required
edges have weight one and the state can be determined using only single-qubit
interference measurements, with no edge-resolution unitaries.
In particular, if $k > 2^n - n$, then $\mathcal{F}_1$ always has a single
connected component; see Theorem~\ref{thm:cnot_limit}~(a).

On the other hand, an $n$-qubit state supported on $(0\cdots0)$ and $(1\cdots1)$
corresponds to a single edge of weight $n$.
Resolving this edge via the ENT implementation requires $n-1$ CNOT gates,
while the PM implementation avoids entangling gates entirely.
The choice between these implementations depends on device noise parameters,
as discussed in Section~\ref{sec:regime_selection}.

In general, for $k \ge 3$ there is always a state $|\psi\ra$ with $k$ nonzero entries requiring 
at least $4(k-1)\lfloor \frac{n}{k} \rfloor$ CNOT gates using our scheme; see Theorem \ref{thm:cnot_limit} (b).

\section{Structural Results and Edge-Resolution Guarantees}
\label{sec:structural_results_and_cnot_guarantees}

In this section, we establish structural results that characterize the
edge-resolution requirements of the sparse tomography algorithm.
These results relate the combinatorial structure of the support of
$|\psi\rangle$ to the number and type of edge-resolution unitaries
required for reconstruction.

Let $n$ be a positive integer and $N=2^n$.
We denote by $\mathcal{G}_n$ the weighted undirected graph with vertex set
$\{0,\dots,N-1\}$, where the weight of an edge $(i,j)$ is given by the
Hamming distance $h(i,j)$ between the binary representations of $i$ and $j$.
That is, if
$i=(i_{n-1}\dots i_0)$ and $j=(j_{n-1}\dots j_0)$, then
$h(i,j)$ is the number of positions at which these two binary strings differ.

For a sparse $n$-qubit state $|\psi\rangle$ with support contained in a
subset of vertices of $\mathcal{G}_n$, the minimal spanning trees (MSTs)
of the induced subgraph determine which amplitude ratios must be resolved. 
The total weight of these trees bounds the number of edge-resolution steps,
and, in the entanglement-assisted implementation, the number of CNOT gates
required by the algorithm. 

\begin{theorem} \label{thm:cnot_limit} Let $n, k$ be a positive integer, and 
$|\psi\ra$ is and $n$-qubit state with $k$ nonzero entries, where
$1 \le k \le  2^n=N$. Suppose $\cG$ is a graph constructed as in Step 1 of the algorithm.
\begin{itemize}
\item[{\rm (a)}] 
Suppose  $k > N-n$. 
Then a minimal spanning tree $T$ of
$\cG$ has weight $k-1$, i.e., every edge of $T$ has weight 1.
In particular, $|\psi\ra$ can be determined by the measurements
of $H_{j_1} |\psi\ra, 
V_{j_1} |\psi\ra,\cdots, H_{j_m} |\psi\ra, V_{j_m} |\psi\ra$ for 
some $j_1, \dots, j_m\in \{0, \dots, n-1\}$
with $m \le k$.
\item[{\rm (b)}] Suppose $3 \le k \le n$
and $n = kq+r$ for some nonnegative integer $q$ and $r \in \{0, \dots, q-1\}$.
Then there is $|\psi\ra$ such that the MST $T$ of $\cG$ has weight 
$2q(k-1) + r$, where $k-2$ of the edges in $T$ have weight 
$2q$ and the remaining edge $T$ has weight $2q+r$.
\end{itemize}
\end{theorem}

\it Proof. \rm (a)  Suppose $\cG$ has $k$ vertices with $m \ge N-n+1$ vertices.
We will prove the assertion by induction on $n$.
We illustrate the ideas for the low dimension cases $n = 2,3,4,5$
to elucidate how the induction is done.

Consider two types of vertices in $\cG$.
Suppose there are $k_1$ so many type 1 vertices of the form $(0i_{n-2} \cdots i_0)$
and $k_2$ so many type 2 vertices of the form  $(1 j_{n-2} \cdots j_0)$, where 
$k_1 + k_2$. We may assume that $k_1 \ge k_2$. Otherwise, we may 
change each vertex $j$ of $\cG$ to $N-1-j$, and prove the result.

Suppose $n=2$ and $\cG$ has $k$ vertices with $k \ge N-n +1 = 3$ vertices.
Then $(k_1,k_2) = (2,1)$ or $(2,2)$.
Then $(00)-(01)$ is a spanning tree the subgraph formed by type 1 vertices.
Every type 2 vertex  is adjacent to a type 1 vertex with an edge of weight 1.
So, for each type 2 vertex, we can have a weight 1 edge to the vertex $(00)$ or $(01)$
to get a spanning tree of weight $k-1$. 

Suppose $n = 3$ and $\cG$ has $m$ vertices with $k \ge N-n+1 = 8-3+1 = 6$ vertices.
Then $(k_1,k_2) = (4,k_2)$ with $k_2 \in \{1,2\}$ or $(3,3)$.
In the former case, there are $k_1-1=3$ edges each of weight 1 connecting
the type 1 vertices. Each type 2 vertex is connected to a type 1 vertex
by an edge of weight 1. So, we can use these edges to form a spanning tree of 
total weights $k-1$.
Suppose $(k_1,k_2) = (3,3)$.
By induction assumption (the case when $n = 2$, 
there are two edges each of weight 1 connecting 
the type $j$ vertices, for $j = 1,2$.  With these 4 edges, we need one more 
edge connecting a type 1 and a type 2 vertex of weight 1 to form 
the desired minimum spanning tree. To this end,
note that there must be a type 1 vertex and type 2 vertex of the form
$i  = (0j_1j_0)$ and $j = (1j_1j_0)$. So, we get the desired additional edge.

Suppose $n = 4$ and $\cG$ has $m$ vertices with $m \ge N -n+1 = 16-4+1 = 13$ vertices.
Then $(k_1,k_2) = (8,k_2)$ with $k_2 \in \{6,7,8\}$ or 
$(k_1,k_2) = (7,k_2)$ with $k_2 \in \{6,7\}$.
In the former case, there are $k_1-1=3$ edges each of weight 1 connecting
the type 1 vertices. Each type 2 vertex is connected to a type 1 vertex
by an edge of weight 1. So, we can use these edges to form a spanning tree of 
total weights $k-1$. In the latter case, 
By induction assumption, 
there are $k_j-1$ edges each of weight 1 connecting 
the type $j$ vertices, for $j = 1,2$.  With these $k_1+k_2-2$ edges, 
we need one more 
edge connecting a type 1 and a type 2 vertex of weight 1 to form 
the desired minimum spanning tree. To this end,
note that there must be a type 1 vertex and type 2 vertex of the form
$i  = (0i_2i_1i_0)$ and $j = (1i_2i_1i_0)$. If it does not hold, then 
type 2 vertices cannot have the form 
$(1i_2 i_1 i_0)$, where  $(0i_2 i_1 i_0)$ is a type 1 vertex.
So, $k_2 \le N/2- k_1$ and $m_1 + m_2 \le N/2$, a contradiction.
Hence, we get the desired additional edge.

Now, assume that we have proved the assertion for subgraph of $\cG_{n-1}$ with $n-1 \ge 4$.
Let $\cG$ be a subgragph of $\cG_n$ with $k_j$ type $j$ vertices, where 
$j = 1,2$ such that $k_1 \ge k_2$ and $k = k_1 +k_2 \ge N-n+1$.
If $k_1 = N/2$, then there are $N/2-1$ edges each of weight 1 connecting
the type 1 vertices. Each type 2 vertex is connected to a type 1 vertex
by an edge of weight 1. So, we can use these edges to form a spanning tree of 
total weights $k-1$. Suppose $k_1 < N/2$.
Then $k_1 \ge k_2 = k-k_1 \ge N-n+1-k_1 \ge N/2-n+2 \ge N/2 - (n-1) + 1$.  
By induction assumption,  there are $k_j-1$ edges each of weight 1 connecting 
the type $j$ vertices, for $j = 1,2$.  With these $k_1+k_2-2$ edges, 
we need one more 
edge connecting a type 1 and a type 2 vertex of weight 1 to form 
the desired minimum spanning tree. To this end,
we claim that there must be a type 1 vertex and type 2 vertex of the form
$i  = (0i_{n-2}\cdots i_1i_0)$ and $j = (1i_{n-2}\cdots i_1i_0)$. 
So, we will have the desired additional edge.

If the claim  does not hold, then 
type 2 vertices cannot have the form 
$(1i_{n-2}\cdots i_1 i_0)$, where  $(0i_{n-2}\cdots i_1 i_0)$ is a type 1 vertex.
So, $k_2 \le N/2- k_1$ and $k_1 + k_2 \le N/2$, a contradiction.

\medskip
(b) Consider the subgraph formed by the binary sequences 
$\ell_1, \dots, \ell_{k}$ such that
$\ell_k$ has the first $(q-1)k$ positions equal 0,
and the rest of the position equal 1; 
for $1 \le j \le k-1$, $\ell_j$ has entries equal to 1 at the positions 
$(j-1)q+1, \dots, jq$, and 0 in other positions.

The $h(\ell_u,\ell_v) = 2q$ for any $1 \le u < v < k$
and $h(\ell_u,\ell_k) = 2q+r$  for any $1 \le u < k$.
To form the minimal spanning tree, we can pick $k-2$ edges each with weight 
$2q$, and then add $\ell_{k}$ to any of the 
vertices $\ell_1, \dots, \ell_{k-1}$. Thus the weight is $2q(k-1)+r$.
\qed

One may also use coding theory to estimate bounds for MST for a subgraph of $\cG_n$;
see for example \cite{vanlintIntroductionCodingTheory1992}.
For example, if $\cG$ is a subgraph of $\cG_n$ with  
$k$ vertices such that every pair of vertices differ
by at least $d$ entries, then the MST will have at least $d(K-1)$.
If any two of the $k$ vertices differ in at least $d$ entries,
then for every vertex $v_j$ we can let $B_j$ be the set of vertices in 
$\cG_n$ differ with $v_j$ in at most  $t = \lfloor (d-1)/2 \rfloor$ positions.
Then each $B_j$ has $\sum_{j=0}^{t} \binom{n}{j}$ elements, and $B_i\cap B_j$
are disjoint if $i \ne j$.
Consequently $k \sum_{j=0}^{t} \binom{n}{j} \le 2^n$.
Hence $k \le 2^n/(\sum_{j=0}^{t} \binom{n}{j})$.
Furthermore, the above bound is known as the Hamming bound and can 
be achieved if $(n,\log_2(k),d) = (2^r-1, 2^r-r-1,3)$. So, there is a graph with 
$k$ vertices with MST having weight $3(k-1)$.
We have the following.

\begin{proposition}
Let $n$ be a positive integer, and let $\cG$ be a subgraph of $\cG_n$ with $k$ vertices.
Suppose  any two of the $k$ vertices differ in at least $d$ entries. Then
$k \le 2^n/(\sum_{j=0}^{t} \binom{n}{j})$. The equality is attainable if  
$(n,\log_2(k),d) = (2^r-1, 2^r-r-1,3)$  so that the MST has weight $3(k-1)$.
\end{proposition}

\subsection{Complexity Guarantees}
\label{ssec:complexity_guarantees}

We analyze the resource complexity of our scheme in terms of the sparsity $k$
(the number of nonzero amplitudes), the system size $n$, and device noise
parameters. The total complexity breaks down into three components:
(i) the number of measurement settings, 
(ii) the number of two-qubit gates used in the preparation unitaries,
and (iii) the number of state copies required for accurate estimation.

\subsubsection{Measurement Settings}
Once the support of $|\psi\rangle$ is identified from an initial
computational-basis measurement, the minimal spanning tree $T$ of the support
graph $\mathcal{G}$ contains $k-1$ edges. For each edge, two measurement
settings (one $H$-type and one $V$-type) are used. Thus the total number of
measurement settings satisfies
\[
M \le 2(k-1) = O(k),
\]
independent of the total dimension $2^n$.

\subsubsection{CNOT gate complexity}
Let $p_j$ denote the number of edges in $T$ of Hamming weight $j$. Then, by
Theorem~\ref{thm:cnot_limit}, the total number of CNOT gates used across all unitaries
$U_1,\dots,U_M$ is
\[
G_{\mathrm{CNOT}}
\;\le\;
2(p_2 + 2p_3 + \cdots + (r-1)p_r)
\;\le\; 
2(r-1)(k-1),
\]
where $r = \max\{ h(\ell_u,\ell_v) : (\ell_u,\ell_v)\in T \}$.
In the worst case, $r = \Theta(n)$, yielding $G_{\mathrm{CNOT}} = O(nk)$,
while in structured or low-diameter supports (e.g., $r=1$),
$G_{\mathrm{CNOT}}=0$.

\subsubsection{Sample Complexity}
Each measurement setting consumes $N_{\mathrm{shots}}$ copies of $|\psi\rangle$,
and the estimation error scales like $O(1/\sqrt{N_{\mathrm{shots}}})$ in the
entanglement-assisted regime.  Thus the total sample complexity satisfies
\[
S = M N_{\mathrm{shots}} = O(k N_{\mathrm{shots}}).
\]
For constant target precision $\epsilon$ and favorable gate fidelities,
choosing $N_{\mathrm{shots}} = O(1/\epsilon^2)$ is sufficient, giving
\[
S = O\!\left(\frac{k}{\epsilon^2}\right),
\]
which is comparable to compressed-sensing optimal bounds for sparse tomography.

\medskip
Altogether, for a $k$-sparse pure $n$-qubit state, the proposed scheme achieves
\[
\boxed{
\text{Measurement settings } = O(k), \qquad
\text{CNOT gates } = O(nk), \qquad
\text{Samples } = O\!\left(\frac{k}{\epsilon^2}\right).
}
\]
Thus our algorithm is efficient for quantum states with $k \ll 2^n$,
and its performance improves further when the support graph admits small
Hamming-weight edges (e.g., chain-like topologies), eliminating or
drastically reducing the need for entangling operations.

\subsubsection{Classical Post-Processing Complexity}
After measurement data are collected, the reconstruction proceeds edge-by-edge
along the minimal spanning tree $T$ of the support graph $\mathcal{G}$.  Each
edge connects two amplitudes $x_i$ and $x_j$ whose ratio is inferred from two
measurement settings (an $H$-type and a $V$-type transformation after the
appropriate edge-resolution unitary).  The classical post-processing for a
single edge therefore consists of:

\begin{enumerate}
    \item extracting empirical frequencies from $N_{\mathrm{shots}}$ samples,
    \item forming the four quadratic equations relating $(x_i,x_j)$ to these statistics,
    \item computing the real and imaginary parts of the target amplitude via linear combinations of the measured interference statistics.
\end{enumerate}

The cost of these operations is $O(N_{\mathrm{shots}})$ to compute empirical
probabilities and $O(1)$ to solve the resulting system.  Since there are $k-1$
edges in $T$, the total classical complexity is
\[
C_{\mathrm{classical}}
    = O(k N_{\mathrm{shots}}) + O(k)
    = O(k N_{\mathrm{shots}}),
\]
matching the sample complexity. No step depends on the full Hilbert-space dimension
$2^n$, and the reconstruction grows linearly with sparsity $k$.

\medskip

The MST itself can be constructed in
\[
O(k \log k + E),
\]
where $E$ is the number of candidate edges considered; using Hamming distance as the
metric, $E$ may be taken as $O(k^2)$ in the worst case, giving
\[
O(k^2).
\]
This computation occurs once, after the support has been identified.  Since
support identification requires only a single computational-basis measurement,
followed by thresholding, the total complexity of structural preprocessing is
polynomial and small compared with data-collection cost.

\medskip

Combining classical and quantum contributions, the overall reconstruction cost is
dominated by $O(kN)$, the same scaling as the total number of samples processed.
Thus, unlike full-state tomography whose post-processing grows as $\Theta(4^n)$
(e.g., via maximum-likelihood estimation or convex optimization), the present scheme
retains polynomial-time decoding even for large $n$ provided the support size $k$
remains small.

\section{Regime Selection}
\label{sec:regime_selection}

Our tomography framework enables two operational modes, depending on
hardware capabilities and the sparsity structure of the unknown state:
(i) an entanglement-assisted scheme, and
(ii) an entanglement-free alternative based on partial mixing using
single-qubit gates. In this section, we analyze salient noise sources of our algorithm, and 
which operational mode to pick. 

\subsection{Noise mechanisms in the entangling and partial-mixing regimes}

The reconstruction of an unknown amplitude pair is affected by four
types of noise: single-qubit gate noise, two-qubit gate noise, readout noise,
and quantum projection noise.  Both schemes ultimately solve the same quadratic
equations, but they differ substantially in how noise enters
the measurement outcomes used in these equations.

\subsubsection{Single-qubit noise}
In the entangling (ENT) scheme, once the CNOTs have aligned the entries $d$ and
$k$ onto a two-dimensional subspace, the interference measurement uses exactly
one non-identity single-qubit gate: either $H$ or $V=HD$.  Thus the
single-qubit noise contribution is of order
\[
    \varepsilon_{1}^{\mathrm{ENT}} \sim p_{\mathrm{1q}},
\]
where $p_{\mathrm{1q}}$ is the single-qubit gate error rate.

In partial mixing (PM), a partial-mixing unitary is applied to each of the
$h$ qubits where $i_d$ and $i_k$ differ.  This requires $h-1$ Hadamards inside
the unitary plus one additional $H$ or $V$ on the pivot qubit, giving
\[
    \varepsilon_{1}^{\mathrm{PM}} \sim h \,p_{\mathrm{1q}}.
\]
PM therefore accumulates single-qubit noise linearly with $h$.

\subsubsection{Two-qubit noise}
The ENT scheme uses $h-1$ CNOT gates to reduce a general Hamming-distance-$h$
pair to the trivial case, producing a systematic distortion
\[
    \varepsilon_{2}^{\mathrm{ENT}}
    \approx (h-1)p_{\mathrm{2q}},
\]
where $p_{\mathrm{2q}}$ is the CNOT error rate.  PM uses no two-qubit gates in its
preparation, replacing the CNOT alignment with a partial-mixing unitary.
Thus PM avoids this error source entirely:
\[
    \varepsilon_{2}^{\mathrm{PM}} = 0.
\]

\subsubsection{Readout noise}
Let $p_{\mathrm{meas}}$ denote the bit-flip probability of the readout channel.
In the entanglement-assisted (ENT) scheme, after the CNOT-based alignment step,
only two computational-basis outcomes corresponding to the aligned amplitude
pair $(i,j)$ contribute to the final quadratic equations.
A readout error symmetrically mixes these two probability masses, reducing the
interference contrast by a factor $(1-2p_{\mathrm{meas}})$.
As a result, the contribution of readout noise to the reconstruction error scales as
\[
    \varepsilon_{\mathrm{meas}}^{\mathrm{ENT}}
    \sim \frac{1}{(1-2p_{\mathrm{meas}})\sqrt{N}},
\]
where $N$ is the number of measurement shots per setting.

In the partial-mixing (PM) scheme, the partial-mixing unitary spreads the two
amplitudes $x_i$ and $x_j$ over $2^{h(i,j)}$ computational-basis states, each of
which enters the quadratic estimator with a different weight.
A readout flip therefore transfers probability mass between bins that contribute
unequally to the estimator, leading to a cumulative error that grows with the
Hamming distance:
\[
    \varepsilon_{\mathrm{meas}}^{\mathrm{PM}} \sim h(i,j)\, p_{\mathrm{meas}}.
\]
Consequently, readout noise remains mild in the ENT scheme but increases
linearly with the Hamming distance in the PM scheme.

\subsubsection{Quantum projection noise}
ENT uses two relevant measurement bins, so projection noise contributes
\[
    \varepsilon_{\mathrm{proj}}^{\mathrm{ENT}} \sim \frac{1}{\sqrt{N}}.
\]

PM distributes the interference over $2^{h}$ bins, each estimated with sampling
error $O(N^{-1/2})$, and these values enter the quadratic estimator.  The
resulting fluctuation scales as
\[
    \varepsilon_{\mathrm{proj}}^{\mathrm{PM}} 
    \sim \sqrt{\frac{2^{h}}{N}}.
\]

\subsection{Crossover behavior}
Combining the leading contributions yields
\[
\varepsilon_{\mathrm{ENT}}(h)
        \approx \sqrt{ [(h-1)\mathrm{2q}]^2 
            + \varepsilon_1^2
            + \frac{1}{(1-2p_{\mathrm{meas}})^2 N} },
\]
\[
\varepsilon_{\mathrm{PM}}(h)
        \approx \sqrt{ (h\,\varepsilon_1)^2
            + (h\,p_{\mathrm{meas}})^2
            + \frac{2^{h}}{N} }.
\]

For small $h$ and large $\mathrm{2q}$, PM is preferred because it avoids CNOT noise.
As $h$ increases, the linear growth of single-qubit and readout errors in PM
eventually becomes dominant.  For sufficiently large Hamming distance, the
projection-noise term $\sqrt{2^{h}/N}$ forces a return to the entangling
scheme, assuming CNOT fidelity remains above a modest threshold.

In summary, partial mixing is advantageous at low $h$ when two-qubit noise is
the limiting factor.  The entangling scheme is preferred when $h$ is large,
when readout noise is significant, or when sampling resources are limited. These are demonstrated in the results of our numerical experiments in Section~\ref{sec:numerical_simulations_and_results}.

\section{Quantum Process Tomography}
\label{sec:qprocess_tomo_closed_system}
A quantum operation, or process,  for a closed system has the form $|\psi\ra \mapsto U|\psi\ra$, where $U$ is unitary in $M_N$. 
One may perform state tomography for $U|j\ra$ for all $j = 0, \dots, N-1$ to determine the columns of $U$.
However, the tomography scheme for a pure state only determines a column up to a unit multiple. 
So, after obtaining the $n$ columns of $U$ up to a unit multiple, we have to determine their relative phases. 
Assume that $U$ is a unitary acting on a $n$-qubit system. 
If $U$ has no zero entries, we may use a $1+2n$ set of measurements to determine each column.
Thus, we need a total of $n(1+2n) + 2n = 2n^2 + 3n$ set of measurements to determine $U$.  
Furthermore, if we put the $N$ columns together to form a matrix $A$, it may not be unitary due to measurement error. 
We need to find the best unitary approximation $V$ of $A$. 
In general, we know that if $A$ has a polar decomposition $PV$, where $V$ is unitary and $P$ is positive semidefinite, then $V$ is the best unitary approximation of $A$.

One may use our scheme to determine $W$ with columns $|w_0\ra, \dots, |x_{n-1}\ra$
by determining the state
$$(I_n \otimes W)|\Psi_n\ra = \frac{1}{\sqrt N}
\begin{pmatrix} |w_0\ra \cr \vdots \cr |w_{N-1}\ra\cr
\end{pmatrix}, \ \ \hbox{ where }  \ \ 
|\Psi_n\ra 
= \sum |i_{n-1}\cdots i_0\ra |i_{n-1} \cdots i_0\ra.$$
Denote by $C(i,j)$ the CNOT gate acting on 
$|q_{2n-1} \cdots q_0\ra$ by changing $q_j$ to $q_i \oplus q_j$, i.e., 
$q_i$ is the control bit and $q_j$ is the target bit.
We have the following scheme for generating $|\Psi_n\ra$.

\medskip\noindent
{\bf A scheme for preparing $|\Psi_n\ra = \sum_{j=1}^n |jj\ra$.}
Let $n$ be a positive integer and $N = 2^n$.
 Then 
 $$|\Psi_n\ra = T_n|0\ra^{\otimes 2n} \quad \hbox{ if } \quad
 T_n = C(2n-1,n-1)\cdots C(n,0) (H^{\otimes n}\otimes I_N).$$

\medskip
One can easily verify the scheme as follows. Note that
$$T_n |0\ra^{\otimes 2n} = C(2n-1,n-1) \cdots C(n,0) \sum |i_{n-1} \cdots i_0\ra |0\ra^{\otimes n}
= \sum |i_{n-1} \cdots i_0\ra |i_{n-1} \cdots i_0\ra = |\Psi_n\ra.$$
The circuit diagrams of $T_1, T_2, T_3$ are shown in the following.

\bigskip
{
\qquad $ \Qcircuit @C=1em @R=1em {
|q_0\ra & & \qw &\qw & \targ & \qw & \qw & \qw
& \qw  & \qw & \qw     \\
|q_1\ra & &  \gate{H} &\qw & \ctrl{-1} & \qw & \qw  & \qw 
& \qw  & \qw & \qw    
}
$
\vskip .3in
\qquad $
\Qcircuit @C=1em @R=1em {
|q_0\ra & & \qw &\qw & \targ & \qw & \qw & \qw & \qw       \\
|q_1\ra & & \qw & \qw & \qw  & \qw & \targ  & \qw &  \qw     \\
|q_2\ra & &  \gate{H} &\qw & \ctrl{-2} & \qw & \qw  & \qw 
& \qw    \\
|q_3\ra & &  \gate{H} &\qw & \qw & \qw & \ctrl{-2}  & \qw 
& \qw     
}
$

\vskip -1.8 in {\hskip 3in 
$ \Qcircuit @C=1em @R=1em {
|q_0\ra & & \qw &\qw & \targ & \qw & \qw & \qw & \qw  & \qw & \qw     \\
|q_1\ra & & \qw & \qw & \qw  & \qw & \targ  & \qw & \qw  & \qw & \qw     \\
|q_2\ra & &  \qw &\qw & \qw  & \qw &  \qw   & \qw 
& \targ   & \qw & \qw    \\
|q_3\ra & &  \gate{H} &\qw & \ctrl{-3}& \qw &  \qw   & \qw 
& \qw  & \qw & \qw    \\
|q_4\ra & &  \gate{H} &\qw & \qw & \qw & \ctrl{-3}   & \qw 
& \qw  & \qw & \qw    \\
|q_5\ra & &  \gate{H} &\qw & \qw  & \qw & \qw  & \qw 
& \ctrl{-3} & \qw & \qw   
}
$
}
}

\section{Numerical Simulations \& Results}
\label{sec:numerical_simulations_and_results}

In this section, we present numerical results obtained using
\texttt{Qiskit}~\cite{javadi-abhariQuantumComputingQiskit2024} on IBM quantum
simulators.
Unless explicitly stated otherwise (as in the noise--parameter sweep of
Fig.~\ref{fig:heatmap}), all experiments use the same noise model derived from
the median reported operation fidelities of the IBM Marrakesh device.
Specifically, we assume fixed single-qubit, two-qubit, and readout error rates,
summarized in Table~\ref{tab:marrakesh_noise}, and identical sampling resources
across all runs.

All experiments were executed on IBM's Aer simulator with
$16{,}384$ shots per measurement setting and repeated $512$ times to obtain
empirical infidelity distributions.
By fixing the physical qubits used in each circuit, we ensure that the Qiskit
transpiler selects a consistent linear layout across the device topology,
eliminating additional SWAP overhead.
For reproducibility, the same random seed is used for all reconstructions.

\begin{table}[ht] 
    \centering
    \begin{tabular}{l c} 
        \hline\hline
        Parameter & Value \\
        \hline
        Single-qubit error $p_{1q}$ & $4.239\times 10^{-4}$ \\
        Two-qubit error $p_{2q}$    & $3.416\times 10^{-3}$ \\
        Readout error $p_{\mathrm{meas}}$ & $1.000 \times 10^{-2}$ \\
        \hline\hline
    \end{tabular}
    \caption{Noise model used}
    \label{tab:marrakesh_noise}
\end{table}

For each run, the reconstruction error was quantified by comparing the predicted
state $\ket{\hat\psi}$ with the target state $\ket{\psi}$ using the fidelity
\(
F(|\hat \psi\rangle\!\langle \hat \psi|,
  |\psi\rangle\!\langle \psi|)
  = |\langle \hat\psi | \psi \rangle|^2.
\)
Note that if $|\langle \hat\psi|\psi\rangle|^2 = \cos^2\theta$, then
$\sqrt{2}\,|\sin\theta|$ equals the Euclidean norm of the matrix
$|\hat\psi\rangle\!\langle\hat\psi| - |\psi\rangle\!\langle\psi|$, providing a
geometric interpretation of the approximation error.  
The infidelity is defined as \(I = 1 - F\).
All preparation circuits are provided in Appendix~\ref{sec:circ-prep-mar}. For state tomography, the epsilon value was chosen to be \(\epsilon = 9.0 \times 10^{-2}\).  For process tomography, the epsilon value was chosen to be \(\epsilon = 9.0 \times 10^{-3}\).  
Code for reproducing the simulations is available at
\href{https://github.com/ypwk/pure-state-tomography}{this repository}.

\begin{figure}[t]
    \centering

    \begin{subfigure}{1\linewidth}
        \centering
        \includegraphics[width=\linewidth]{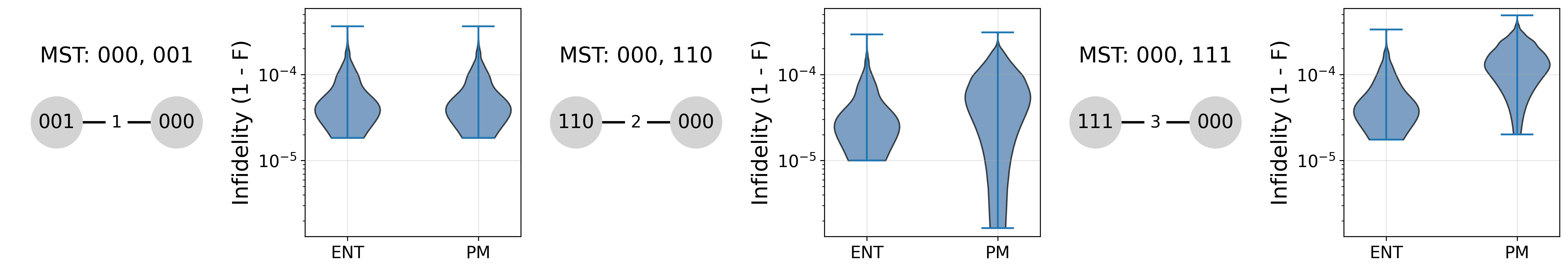}
        \caption{Two–support states}
        \label{fig:support2}
    \end{subfigure}

    \vspace{1em}

    \begin{subfigure}{1\linewidth}
        \centering
        \includegraphics[width=\linewidth]{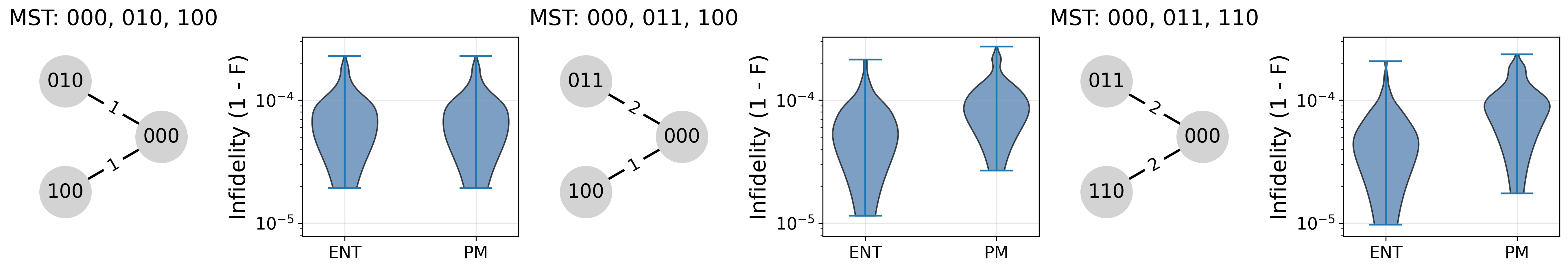}
        \caption{Three–support states}
        \label{fig:support3}
    \end{subfigure}

    \caption{
        Comparison of entanglement-assisted (ENT) and partial-mixing (PM)
        tomography across different support sizes.  
        Each subfigure shows the minimum-spanning-tree structure of the
        nonzero amplitudes alongside the corresponding infidelity
        distributions.  
    }
    \label{fig:support_comparison}
\end{figure}

\subsection{Three-Qubit States}

We examine several three-qubit examples, each defined by a small support on the
computational basis together with the minimum-spanning tree (MST) over those
basis states.  The MST encodes the effective “geometry’’ of the amplitudes:
trivial MSTs correspond to two basis states at Hamming distance one, while
larger supports introduce additional nodes and longer edges. Although these 
examples are small, they are useful diagnostic cases: the MST
structure is simple enough to inspect visually, yet the differences in mixing
dimension are large enough for PM’s scaling behavior to emerge clearly.

Figure~\ref{fig:support_comparison} shows the corresponding MSTs alongside the
ENT and PM infidelity distributions.  For states with trivial MST structure
(i.e., only weight-1 edges), the two protocols produce nearly identical results.
This is expected: the circuits for ENT and PM coincide in this special case,
and identical random seeds yield matching sampling fluctuations.  Once the
support grows, the distributions begin to separate, reflecting the increased
mixing dimension experienced by the PM method.

\begin{figure}[!t]
    \centering
    \includegraphics[width=\linewidth]{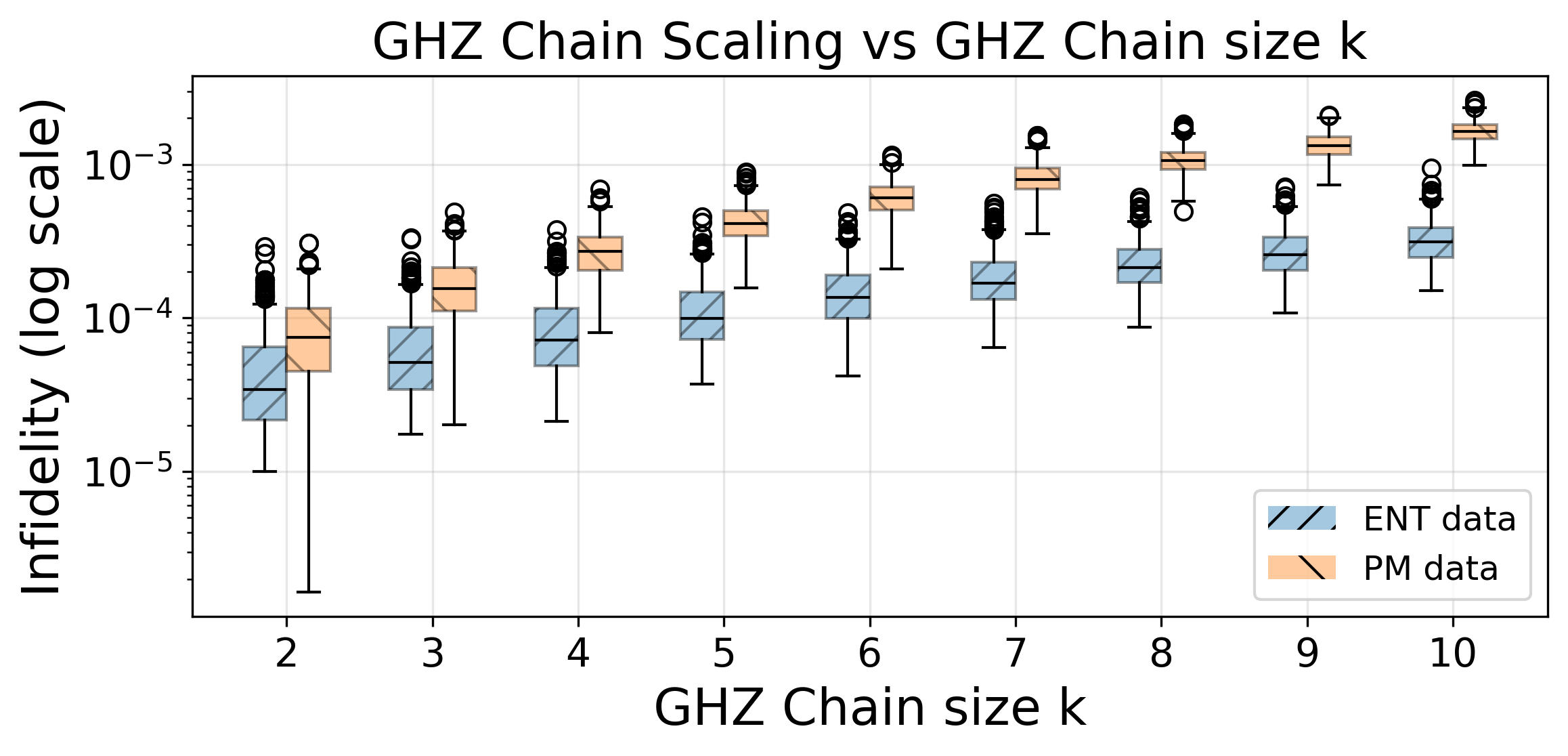}
    \caption{log-infidelity distributions for GHZ chains of size $k=2$--$10$ under 
        the IBM Marrakesh noise model, comparing entangling (blue) and partial-mixing (orange) protocols.}
    \label{fig:entanglement_classes}
\end{figure}

\subsection{GHZ States}

To assess protocol performance across increasing multipartite entanglement, we
simulate GHZ states of size $n=2$--$10$ under a realistic device-level noise
model. Noise parameters are taken from median operation fidelities reported for
IBM Marrakesh, a 156-qubit superconducting processor. Synthetic measurement
data are generated using these parameters and reconstructed using both
tomography protocols. To ensure that differences between the two schemes 
are not confounded by routing overhead, we assume all-to-all qubit connectivity, 
eliminating any contribution from SWAP operations.

Figure~\ref{fig:entanglement_classes} shows the resulting log-infidelity
distributions for both methods as box-and-whisker plots for each chain size \(k\). 
The noise model used for these simulations are summarized in the table below.

Note that in the regime described by the above table, the partial mixing method suffers from the model's high readout error, due to increased uncertainty in its distributed probability mass. Therefore, the entangling method wins. 

\begin{figure}[h]
    \centering
    \includegraphics[width=0.9\linewidth]{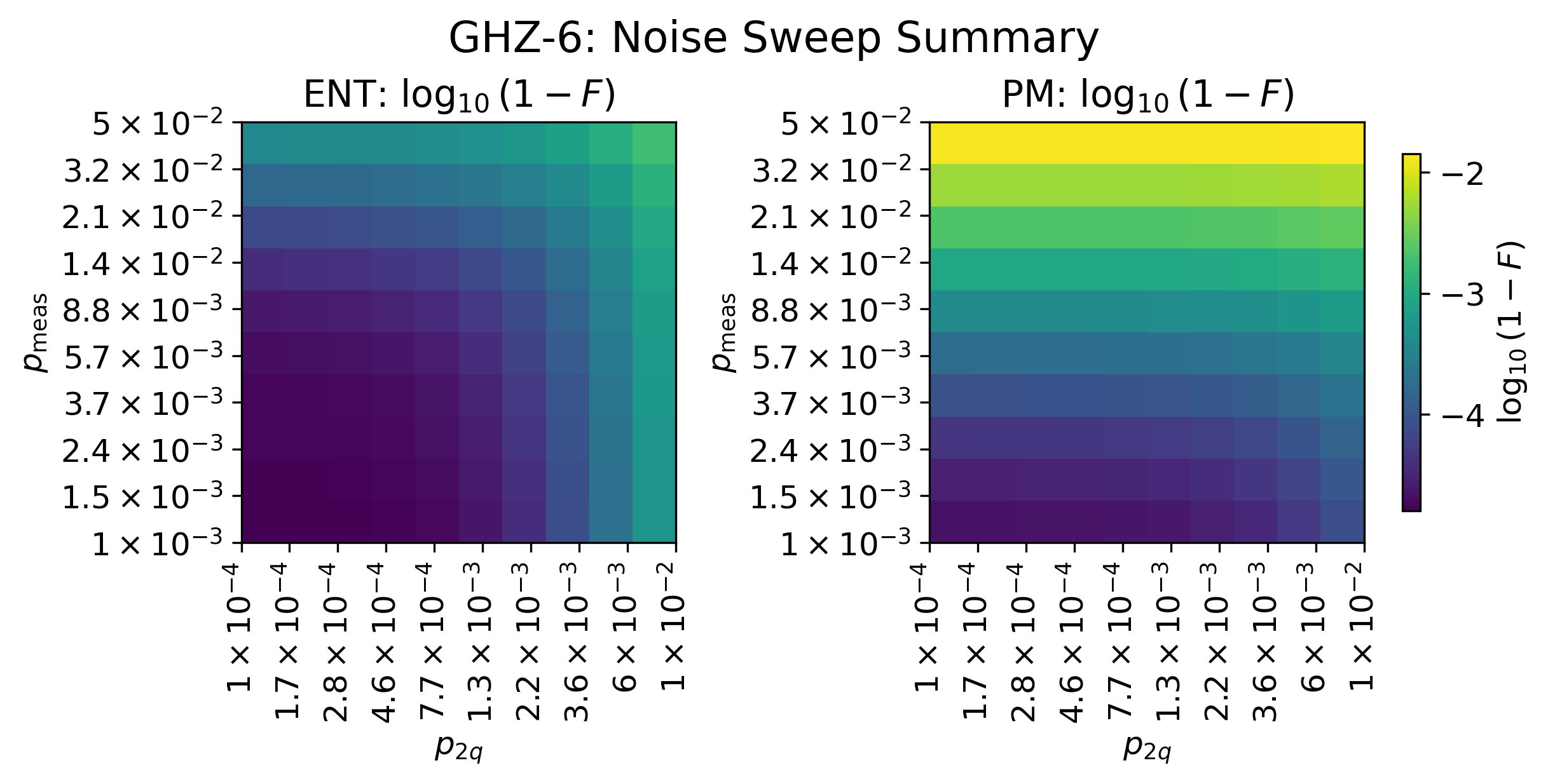}
    \caption{Noise–parameter sweep for a GHZ-6 state.
    Each panel shows reconstruction log-infidelity as a function of the
    two-qubit depolarizing rate $p_{2q}$ and measurement error $p_{\mathrm{meas}}$.
    The heatmaps illustrate the sensitivity of the entangling and partial-mixing 
    tomography protocols to gate and readout noise, revealing the parameter regimes 
    in which each method breaks down.}
    \label{fig:heatmap}
\end{figure}

Figure~\ref{fig:heatmap} illustrates how GHZ-6 reconstruction responds to
variations in the two-qubit depolarizing rate $p_{2q}$ and the readout error
$p_{\mathrm{meas}}$.  The partial-mixing method shows strong dependence on
$p_{\mathrm{meas}}$, reflecting its shot-noise scaling: when amplitudes spread
over many basis states, each outcome carries less weight, and statistical noise
dominates. The entanglement-assisted (ENT) scheme maintains lower infidelity across the
same region because its CNOT-based edge-resolution step concentrates
probability mass onto a two-dimensional subspace before interference,
thereby suppressing the amplification of projection and readout noise.

These trends match the regime analysis in Section~\ref{sec:regime_selection}.
When two-qubit errors are below the threshold set by 
$(h(i,j)-1)e_2 \ll 1/\sqrt{N}$, the MST protocol remains stable.  
Partial mixing, on the other hand, is limited by 
$\sqrt{2^{h(i,j)}/N}$ and becomes noise-dominated once readout errors or
amplitude spreading increase.  The GHZ-6 heatmap therefore marks the point at
which the system transitions from a CNOT-limited regime to a shot-noise-limited
one.

\subsection{Process Tomography}
\label{ssec:process_tomography}

\begin{figure}[t]
    \centering

    \begin{subfigure}{0.48\linewidth}
        \centering
        \begin{minipage}[c]{\linewidth} 
            \centering
            $
            \Qcircuit @C=1em @R=1em {
            |q_0\rangle && \qw & \qw & \targ & \qw & \qw & \qw & \gate{R_y} & \qw & \targ & \qw \\
            |q_1\rangle && \qw & \qw & \qw   & \qw & \targ & \qw & \gate{R_x} & \qw & \ctrl{-1} & \qw \\
            |q_2\rangle && \gate{H} & \qw & \ctrl{-2} & \qw & \qw & \qw & \qw & \qw & \qw & \qw \\
            |q_3\rangle && \gate{H} & \qw & \qw       & \qw & \ctrl{-2} & \qw & \qw & \qw & \qw & \qw
            }
            $
            \caption{Process-tomography circuit.}
        \end{minipage}
    \end{subfigure}
    \hfill
    \begin{subfigure}{0.48\linewidth}
        \centering
        \begin{minipage}[c]{\linewidth} 
            \centering
            \includegraphics[width=\linewidth]{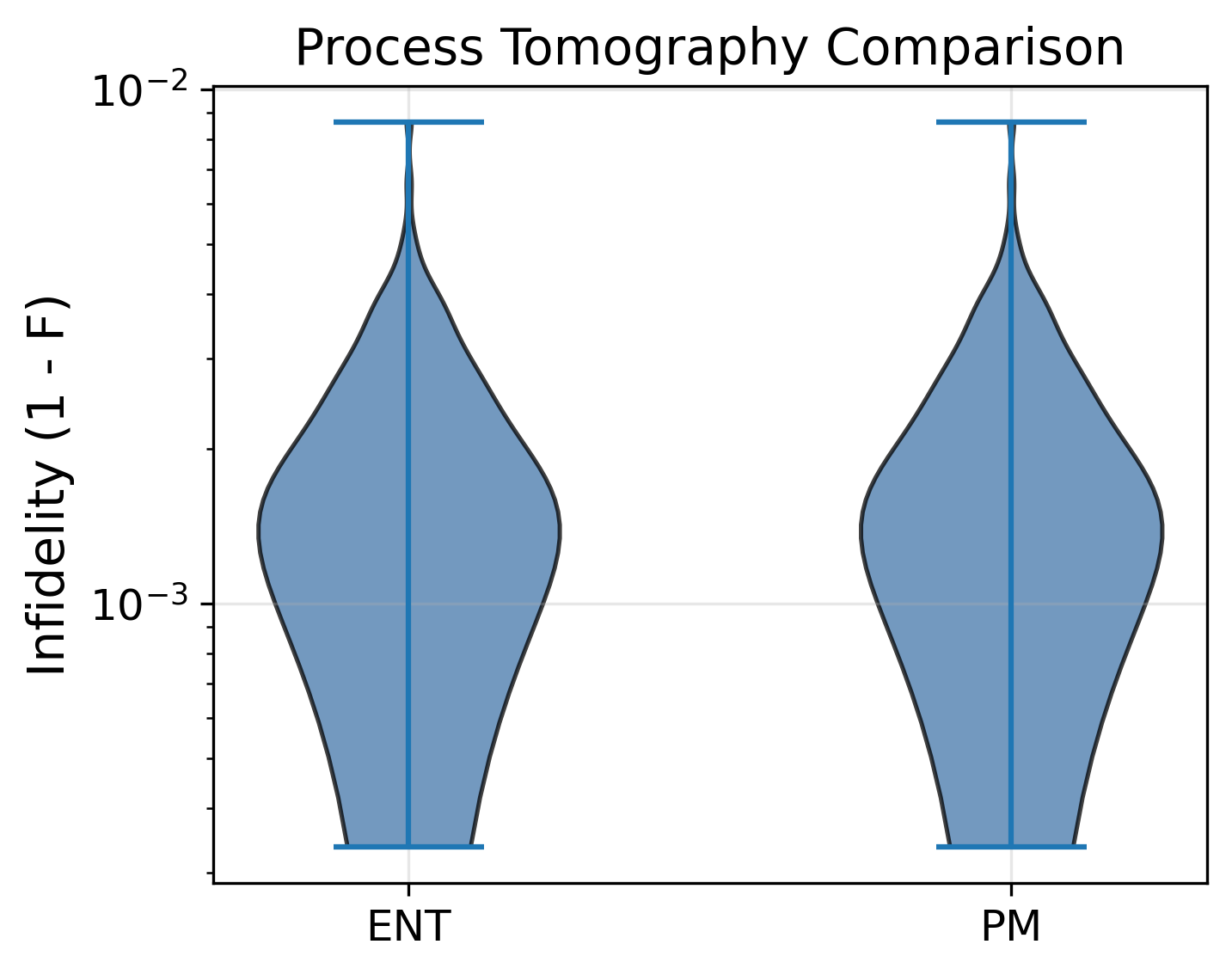}
            \caption{PM vs.\ ENT comparison.}
        \end{minipage}
    \end{subfigure}

    \caption{Two-qubit process tomography (see Sec.~\ref{ssec:process_tomography}).}
    \label{fig:process_combined}
\end{figure}

In addition to pure-state tomography, we evaluated a representative example of
two-qubit process tomography.  
The target operation was the unitary
\(
W_1 = \mathrm{CNOT}\, R_x(\theta) R_y(\phi)
\)
with \((\theta,\phi) = (\pi/4,\pi/3)\), implemented as shown in
Fig.~\ref{fig:process_combined}.
To characterize \(W_1\), we prepared the state
\(
|\tilde v\rangle = (I_4 \otimes W_1)\,|\Psi_2\rangle,
\quad
|\Psi_2\rangle = 2^{-1}\!\sum_j |jj\rangle,
\)
and performed standard state tomography on \(|\tilde v\rangle\).
Both the MST-assisted and partial-mixing protocols recovered the process with
high fidelity, with MST exhibiting slightly reduced variance across repetitions.
This example confirms that the methods apply directly to general CPTP process
estimation in addition to pure-state reconstruction.

\medskip

Overall, the numerical simulations indicate that the entanglement-assisted (ENT)
protocol provides consistently high reconstruction fidelity with narrow
variance across all structural classes of states tested, including those with
large Hamming distance, non-stabilizer amplitude structure, and increasing
system size.
The partial-mixing (PM) protocol performs well for low Hamming-distance edges
and simple superpositions, while displaying increased variance for larger and
more structurally complex supports due to amplitude spreading.
Together, these results highlight the complementary strengths of the two
protocols: both exploit the MST structure of the support to minimize the number
of interference steps, while differing in how circuit-level noise is traded
between entangling operations and quantum projection noise.

\section{Discussion and Conclusion}
\label{sec:discussion_and_conclusion}

The results presented above demonstrate that prior structural information about a
pure quantum state—specifically, sparsity in the computational basis—can be
systematically exploited to design tomographic protocols whose resource
requirements scale with the sparsity parameter $k$, rather than with the full
Hilbert space dimension $2^n$. To our knowledge, this is the first tomographic framework that converts
support geometry directly into circuit-level resource bounds via an explicit
graph-theoretic construction.
Once the support of $|\psi\rangle$ has been identified, the reconstruction task
reduces to determining the relative phases and magnitudes of $k$ nonzero
amplitudes.
The graph-theoretic formulation in terms of a weighted support graph
$\mathcal{G}$ and its minimum spanning tree (MST) provides a unifying organizing
principle: each edge of the MST corresponds to a two-dimensional interference
problem, while the total tomographic cost is governed by the edge weights, i.e.,
the relevant Hamming distances.
In this way, the combinatorial structure of the support directly controls the
number of measurement settings, the depth of entangling circuits, and the
dominant noise mechanisms affecting reconstruction fidelity.

From a complexity perspective, the proposed scheme achieves $O(k)$ distinct
measurement settings, $O(nk)$ two-qubit gates in the worst case, and a sample
complexity $O(k/\epsilon^2)$ to achieve reconstruction error $\epsilon$ under
standard statistical assumptions.
This scaling compares favorably with conventional informationally complete
tomography on $n$ qubits, which typically requires $\Omega(4^n)$ measurement
settings for full density-matrix reconstruction
\cite{altepeterPhotonicStateTomography2005,jamesMeasurementQubits2001}.
Relative to compressed-sensing approaches for low-rank quantum states
\cite{cramerEfficientQuantumState2010,flammiaQuantumTomographyCompressed2012},
the present framework trades worst-case instance optimality for explicit,
circuit-level control over entangling depth and measurement structure.
This distinction is particularly relevant in near-term devices, where
two-qubit gate errors, connectivity constraints, and readout noise often dominate
over classical post-processing costs.
Moreover, the structural result of Theorem~\ref{thm:cnot_limit} identifies a
regime in which no entangling operations are required at all: whenever the MST
edges all have unit weight, the reconstruction can be performed using only
single-qubit gates and measurements.

The numerical simulations performed using IBM’s Aer simulator with
Marrakesh-derived median noise parameters further clarify the practical behavior
of the two operational regimes introduced in
Section~\ref{sec:regime_selection}.
Across all benchmark families considered—including two-component
superpositions with varying Hamming distance, multi-component sparse states,
non-stabilizer amplitude distributions, and GHZ chains of increasing size—the
entanglement-assisted (ENT) protocol consistently yields high-fidelity
reconstructions with narrow variance.
This behavior is consistent with the theoretical analysis: by concentrating
probability mass onto a two-dimensional subspace prior to interference, the ENT
protocol suppresses the amplification of quantum projection noise and remains
robust provided that accumulated two-qubit gate errors remain below the
effective crossover threshold.
In contrast, the partial-mixing (PM) protocol exhibits a clear dependence on the
Hamming distance between support components.
While PM performs comparably to ENT for low-weight superpositions, its fidelity
distribution broadens as probability mass is coherently spread over larger
subspaces, in agreement with the
$\varepsilon_{\mathrm{PM}} \sim \sqrt{2^{h(i,j)}/N}$ scaling derived in
Section~\ref{sec:regime_selection}.
The noise-parameter sweep for the GHZ-6 state highlights this transition
explicitly, delineating the boundary between CNOT-limited and shot-noise-limited
operating regimes.

The process-tomography example illustrates that the sparse-entry framework
extends naturally beyond pure-state reconstruction.
By preparing a maximally entangled input state and performing state tomography
on $(I \otimes U)|\Psi_n\rangle$, the same MST-guided interference strategy can be
used to recover the columns of an unknown unitary operation.
Although the numerical study presented here focuses on a two-qubit gate
sequence, the construction of
Section~\ref{sec:qprocess_tomo_closed_system} applies directly to arbitrary
$n$-qubit unitaries, with the resulting estimate projected onto the space of
physical channels via polar decomposition.
This suggests that sparse-entry optimization may serve as a useful primitive for
characterizing coherent dynamics when the relevant columns of the process matrix
exhibit sparsity or related structure.

Several limitations and open directions remain.
First, the present analysis assumes exact sparsity and a reliable procedure for
support identification via thresholding in the computational basis.
In experimental settings, small but non-negligible leakage amplitudes may lead to
effective approximate sparsity, and a more refined statistical treatment would be
required to quantify robustness under such deviations.
Second, the resource analysis focuses primarily on measurement settings and
entangling gate counts; incorporating device-specific constraints such as
restricted connectivity, time-correlated noise, or asymmetric readout errors
could further refine the regime-selection criteria and motivate
hardware-aware variants of the MST construction.
Third, while the minimum spanning tree (MST) provides an instance-dependent
organization of the reconstruction task, a systematic characterization of
worst-case support geometries remains open.
In particular, it would be valuable to identify or construct families of
$k$-sparse supports whose MSTs maximize the total edge weight or depth, thereby
provably saturating the $O(nk)$ two-qubit gate bound.
Such worst-case examples would sharpen the theoretical guarantees of the method
and clarify how often unfavorable geometries arise among random or structured
supports.
Finally, while the numerical benchmarks span a range of structured state
families, it would be informative to compare performance against other modern
approaches, including adaptive Bayesian estimators
\cite{blumekohoutOptimalReliableEstimation2010}, classical-shadow-based methods
\cite{flammiaDirectFidelityEstimation2011}, and neural-network-assisted
tomography \cite{ahmedQuantumStateTomography2021,weiNeuralshadowQuantumState2024},
under matched experimental assumptions.

In conclusion, we have introduced a sparse-entry tomography framework that
leverages the combinatorial structure of a quantum state’s support to achieve
reconstruction costs scaling linearly with the sparsity parameter $k$.
By organizing the reconstruction problem around the minimum spanning tree of the
support graph, the protocol provides explicit, circuit-level control over the
tradeoff between entangling operations and statistical noise.
Theoretical analysis identifies sharp operating regimes for entanglement-assisted
and entanglement-free implementations, while numerical simulations using
realistic noise models validate these predictions across a diverse set of state
families and a representative process-tomography instance.
Together, these results suggest that structural priors such as sparsity can be
used to design tomographic protocols that are both theoretically efficient and
well matched to the constraints of near-term quantum hardware.

\bigskip
\noindent
{\bf \large Acknowledgment}

Li is an affiliate member
of the Institute for Quantum Computing of the University of Waterloo, 
and also an affiliate member of the Quantum Science \& Engineering Center of the George Mason University; his research
was partially supported by the Simons Foundation Grant 851334. 
We would like to thank Professors Mikio Nakahara
and Gexin Yu for helpful discussions.

\printbibliography

@article{ahmedQuantumStateTomography2021,
  title = {Quantum {{State Tomography}} with {{Conditional Generative Adversarial Networks}}},
  author = {Ahmed, Shahnawaz and Sánchez Muñoz, Carlos and Nori, Franco and Kockum, Anton Frisk},
  date = {2021-09-27},
  journaltitle = {Physical Review Letters},
  shortjournal = {Phys. Rev. Lett.},
  volume = {127},
  number = {14},
  pages = {140502},
  publisher = {American Physical Society},
  doi = {10.1103/PhysRevLett.127.140502},
  url = {https://link.aps.org/doi/10.1103/PhysRevLett.127.140502},
  urldate = {2025-02-17}
}

@incollection{altepeterPhotonicStateTomography2005,
  title = {Photonic {{State Tomography}}},
  booktitle = {Advances {{In Atomic}}, {{Molecular}}, and {{Optical Physics}}},
  author = {Altepeter, J. B. and Jeffrey, E. R. and Kwiat, P. G.},
  editor = {Berman, P. R. and Lin, C. C.},
  date = {2005-01-01},
  volume = {52},
  pages = {105--159},
  publisher = {Academic Press},
  doi = {10.1016/S1049-250X(05)52003-2},
  url = {https://www.sciencedirect.com/science/article/pii/S1049250X05520032},
  urldate = {2025-07-01}
}

@online{bellanteQuantumSparseRecovery2025,
  title = {Quantum {{Sparse Recovery}} and {{Quantum Orthogonal Matching Pursuit}}},
  author = {Bellante, Armando and Vanerio, Stefano and Zanero, Stefano},
  date = {2025-10-08},
  eprint = {2510.06925},
  eprinttype = {arXiv},
  eprintclass = {quant-ph},
  doi = {10.48550/arXiv.2510.06925},
  url = {http://arxiv.org/abs/2510.06925},
  urldate = {2025-12-18},
  pubstate = {prepublished}
}

@article{blumekohoutOptimalReliableEstimation2010,
  title = {Optimal, Reliable Estimation of Quantum States},
  author = {Blume-Kohout, Robin},
  date = {2010-04-20},
  journaltitle = {New Journal of Physics},
  shortjournal = {New J. Phys.},
  volume = {12},
  number = {4},
  pages = {043034},
  issn = {1367-2630},
  doi = {10.1088/1367-2630/12/4/043034},
  url = {https://iopscience.iop.org/article/10.1088/1367-2630/12/4/043034},
  urldate = {2025-02-20},
  langid = {english}
}

@article{childsQuantumInformationPrecision2000,
  title = {Quantum Information and Precision Measurement},
  author = {Childs, Andrew M. and Preskill, John and Renes, Joseph},
  date = {2000-02-01},
  journaltitle = {Journal of Modern Optics},
  volume = {47},
  number = {2--3},
  pages = {155--176},
  publisher = {Taylor \& Francis},
  issn = {0950-0340},
  doi = {10.1080/09500340008244034},
  url = {https://www.tandfonline.com/doi/abs/10.1080/09500340008244034},
  urldate = {2025-07-01}
}

@article{cramerEfficientQuantumState2010,
  title = {Efficient Quantum State Tomography},
  author = {Cramer, Marcus and Plenio, Martin B. and Flammia, Steven T. and Somma, Rolando and Gross, David and Bartlett, Stephen D. and Landon-Cardinal, Olivier and Poulin, David and Liu, Yi-Kai},
  date = {2010-12-21},
  journaltitle = {Nature Communications},
  shortjournal = {Nat Commun},
  volume = {1},
  number = {1},
  pages = {149},
  publisher = {Nature Publishing Group},
  issn = {2041-1723},
  doi = {10.1038/ncomms1147},
  url = {https://www.nature.com/articles/ncomms1147},
  urldate = {2025-02-17},
  langid = {english}
}

@article{emersonScalableNoiseEstimation2005,
  title = {Scalable Noise Estimation with Random Unitary Operators},
  author = {Emerson, Joseph and Alicki, Robert and Życzkowski, Karol},
  date = {2005-09},
  journaltitle = {Journal of Optics B: Quantum and Semiclassical Optics},
  shortjournal = {J. Opt. B: Quantum Semiclass. Opt.},
  volume = {7},
  number = {10},
  pages = {S347},
  issn = {1464-4266},
  doi = {10.1088/1464-4266/7/10/021},
  url = {https://dx.doi.org/10.1088/1464-4266/7/10/021},
  urldate = {2025-07-01},
  langid = {english}
}

@article{flammiaDirectFidelityEstimation2011,
  title = {Direct {{Fidelity Estimation}} from {{Few Pauli Measurements}}},
  author = {Flammia, Steven T. and Liu, Yi-Kai},
  date = {2011-06-08},
  journaltitle = {Physical Review Letters},
  shortjournal = {Phys. Rev. Lett.},
  volume = {106},
  number = {23},
  pages = {230501},
  publisher = {American Physical Society},
  doi = {10.1103/PhysRevLett.106.230501},
  url = {https://link.aps.org/doi/10.1103/PhysRevLett.106.230501},
  urldate = {2025-07-08}
}

@article{flammiaQuantumTomographyCompressed2012,
  title = {Quantum Tomography via Compressed Sensing: Error Bounds, Sample Complexity and Efficient Estimators},
  shorttitle = {Quantum Tomography via Compressed Sensing},
  author = {Flammia, Steven T and Gross, David and Liu, Yi-Kai and Eisert, Jens},
  date = {2012-09},
  journaltitle = {New Journal of Physics},
  shortjournal = {New J. Phys.},
  volume = {14},
  number = {9},
  pages = {095022},
  publisher = {IOP Publishing},
  issn = {1367-2630},
  doi = {10.1088/1367-2630/14/9/095022},
  url = {https://dx.doi.org/10.1088/1367-2630/14/9/095022},
  urldate = {2025-07-01},
  langid = {english}
}

@article{hradilQuantumstateEstimation1997,
  title = {Quantum-State Estimation},
  author = {Hradil, Z.},
  date = {1997-03-01},
  journaltitle = {Physical Review A},
  shortjournal = {Phys. Rev. A},
  volume = {55},
  number = {3},
  pages = {R1561-R1564},
  publisher = {American Physical Society},
  doi = {10.1103/PhysRevA.55.R1561},
  url = {https://link.aps.org/doi/10.1103/PhysRevA.55.R1561},
  urldate = {2025-07-01}
}

@article{jamesMeasurementQubits2001,
  title = {Measurement of Qubits},
  author = {James, Daniel F. V. and Kwiat, Paul G. and Munro, William J. and White, Andrew G.},
  date = {2001-10-16},
  journaltitle = {Physical Review A},
  shortjournal = {Phys. Rev. A},
  volume = {64},
  number = {5},
  pages = {052312},
  publisher = {American Physical Society},
  doi = {10.1103/PhysRevA.64.052312},
  url = {https://link.aps.org/doi/10.1103/PhysRevA.64.052312},
  urldate = {2025-07-01}
}

@online{javadi-abhariQuantumComputingQiskit2024,
  title = {Quantum Computing with {{Qiskit}}},
  author = {Javadi-Abhari, Ali and Treinish, Matthew and Krsulich, Kevin and Wood, Christopher J. and Lishman, Jake and Gacon, Julien and Martiel, Simon and Nation, Paul D. and Bishop, Lev S. and Cross, Andrew W. and Johnson, Blake R. and Gambetta, Jay M.},
  date = {2024-06-19},
  eprint = {2405.08810},
  eprinttype = {arXiv},
  eprintclass = {quant-ph},
  doi = {10.48550/arXiv.2405.08810},
  url = {http://arxiv.org/abs/2405.08810},
  urldate = {2025-07-01},
  pubstate = {prepublished}
}

@article{kieuTravellingSalesmanProblem2019,
  title = {The Travelling Salesman Problem and Adiabatic Quantum Computation: An Algorithm},
  shorttitle = {The Travelling Salesman Problem and Adiabatic Quantum Computation},
  author = {Kieu, Tien D.},
  date = {2019-02-10},
  journaltitle = {Quantum Information Processing},
  shortjournal = {Quantum Inf Process},
  volume = {18},
  number = {3},
  pages = {90},
  issn = {1573-1332},
  doi = {10.1007/s11128-019-2206-9},
  url = {https://doi.org/10.1007/s11128-019-2206-9},
  urldate = {2025-12-18},
  langid = {english}
}

@article{kruskalShortestSpanningSubtree,
  title = {On the {{Shortest Spanning Subtree}} of a {{Graph}} and the {{Traveling Salesman Problem}}},
  author = {Kruskal, Joseph B},
  langid = {english}
}

@book{nielsenQuantumComputationQuantum2010,
  title = {Quantum {{Computation}} and {{Quantum Information}}: 10th {{Anniversary Edition}}},
  shorttitle = {Quantum {{Computation}} and {{Quantum Information}}},
  author = {Nielsen, Michael A. and Chuang, Isaac L.},
  date = {2010-12-08},
  publisher = {Cambridge University Press},
  doi = {10.1017/CBO9780511976667},
  url = {https://www.cambridge.org/highereducation/books/quantum-computation-and-quantum-information/01E10196D0A682A6AEFFEA52D53BE9AE},
  urldate = {2025-07-01},
  isbn = {9780511976667},
  langid = {english},
  organization = {Higher Education from Cambridge University Press}
}

@article{padmasolaSolvingTravelingSalesman2025,
  title = {Solving the Traveling Salesman Problem via Different Quantum Computing Architectures},
  author = {Padmasola, Venkat Aneesh and Li, Zhaotong and Chatterjee, Rupak and Dyk, Wesley},
  date = {2025-11-07},
  journaltitle = {International Journal of Quantum Information},
  publisher = {World Scientific Publishing Company},
  doi = {10.1142/S0219749925400039},
  url = {https://www.worldscientific.com/worldscinet/ijqi},
  urldate = {2025-12-18},
  langid = {english}
}

@article{primShortestConnectionNetworks1957,
  title = {Shortest Connection Networks and Some Generalizations},
  author = {Prim, R. C.},
  date = {1957-11},
  journaltitle = {The Bell System Technical Journal},
  volume = {36},
  number = {6},
  pages = {1389--1401},
  issn = {0005-8580},
  doi = {10.1002/j.1538-7305.1957.tb01515.x},
  url = {https://ieeexplore.ieee.org/document/6773228},
  urldate = {2025-07-01}
}

@online{salesAdiabaticQuantumComputing2023,
  title = {Adiabatic {{Quantum Computing}} for {{Logistic Transport Optimization}}},
  author = {Sales, Juan Francisco Ariño and Araos, Raúl Andres Palacios},
  date = {2023-01-18},
  eprint = {2301.07691},
  eprinttype = {arXiv},
  eprintclass = {quant-ph},
  doi = {10.48550/arXiv.2301.07691},
  url = {http://arxiv.org/abs/2301.07691},
  urldate = {2025-12-18},
  langid = {english},
  pubstate = {prepublished}
}

@article{schreiberTomographyParametrizedQuantum2025,
  title = {Tomography of {{Parametrized Quantum States}}},
  author = {Schreiber, Franz J. and Eisert, Jens and Meyer, Johannes Jakob},
  date = {2025-06-06},
  journaltitle = {PRX Quantum},
  shortjournal = {PRX Quantum},
  volume = {6},
  number = {2},
  pages = {020346},
  publisher = {American Physical Society},
  doi = {10.1103/PRXQuantum.6.020346},
  url = {https://link.aps.org/doi/10.1103/PRXQuantum.6.020346},
  urldate = {2025-12-18}
}

@book{vanlintIntroductionCodingTheory1992,
  title = {Introduction to {{Coding Theory}}},
  author = {Van Lint, J. H.},
  date = {1992},
  series = {Graduate {{Texts}} in {{Mathematics}}},
  volume = {86},
  publisher = {Springer},
  location = {Berlin, Heidelberg},
  doi = {10.1007/978-3-662-00174-5},
  url = {http://link.springer.com/10.1007/978-3-662-00174-5},
  urldate = {2025-07-01},
  isbn = {978-3-662-00176-9}
}

@article{weiNeuralshadowQuantumState2024,
  title = {Neural-Shadow Quantum State Tomography},
  author = {Wei, Victor and Coish, W. A. and Ronagh, Pooya and Muschik, Christine A.},
  date = {2024-06-06},
  journaltitle = {Physical Review Research},
  shortjournal = {Phys. Rev. Res.},
  volume = {6},
  number = {2},
  pages = {023250},
  publisher = {American Physical Society},
  doi = {10.1103/PhysRevResearch.6.023250},
  url = {https://link.aps.org/doi/10.1103/PhysRevResearch.6.023250},
  urldate = {2025-02-17}
}

@article{xieReconstructingQuantumStates2023,
  title = {Reconstructing {{Quantum States}} from {{Sparse Measurements}}},
  author = {Xie, Yufan and Guo, Chu and Peng, Zhihui},
  date = {2023-01},
  journaltitle = {Electronics},
  volume = {12},
  number = {5},
  pages = {1096},
  publisher = {Multidisciplinary Digital Publishing Institute},
  issn = {2079-9292},
  doi = {10.3390/electronics12051096},
  url = {https://www.mdpi.com/2079-9292/12/5/1096},
  urldate = {2025-02-17},
  issue = {5},
  langid = {english}
}

\newpage 
\appendix
\addcontentsline{toc}{section}{Appendix}
\section*{Appendix}

\section{Experimental Configurations}
\subsection{Circuit Diagrams for MST and Randomized State Tomography}
\label{sec:circ-prep-mar}

This subsection presents the circuit constructions used to prepare the benchmark
states employed in the numerical simulations comparing the minimum--spanning--tree
(MST) protocol and the randomized state tomography protocol.  
All three-qubit states considered here take the form
\(
\ket{\psi} = (x_{000}, x_{001}, x_{010}, \ldots, x_{111})^{t},
\)
where only a small subset of amplitudes is nonzero.  
For clarity, the preparation circuits are grouped according to the number of
nonzero computational-basis components.

\subsubsection*{Two-component superpositions}

Figure~\ref{fig:support2} illustrates representative circuits that
prepare superpositions supported on exactly two basis states.  These include
superpositions differing in one, two, or three bit positions.  For example,
the preparations for the cases $(x_{000}, x_{001})$, $(x_{000}, x_{011})$, and
$(x_{000}, x_{111})$ are:
\begin{enumerate}
    \item Case $\{x_{000}, x_{001}\}$:
    $$\Qcircuit @C=1em @R=1em {
        |q_0\ra & & \qw         & \qw \\
        |q_1\ra & & \qw         & \qw \\
        |q_2\ra & & \gate{H}    & \qw \\   
    }$$

    \item Case $\{x_{000}, x_{011}\}$:
    $$\Qcircuit @C=1em @R=1em {
        |q_0\ra & & \qw      & \qw      & \qw       & \qw & \qw \\
        |q_1\ra & & \qw      & \qw      & \targ     & \qw & \qw \\
        |q_2\ra & & \gate{H} & \gate{X} & \ctrl{-1} & \qw & \qw \\   
    }$$

    \item Case $\{x_{000}, x_{111}\}$:
    $$\Qcircuit @C=1em @R=1em {
        |q_0\ra & & \qw      & \qw      & \qw       & \targ     & \qw \\
        |q_1\ra & & \qw      & \qw      & \targ     & \qw       & \qw \\
        |q_2\ra & & \gate{H} & \gate{X} & \ctrl{-1} & \ctrl{-2} & \qw \\
    }$$
\end{enumerate}

Two-component states that differ on all three qubits (e.g., between
$011$ and $100$, or between $110$ and $001$) are prepared analogously:
\begin{enumerate}
    \item Case $\{x_{011}, x_{100}\}$:
    $$\Qcircuit @C=1em @R=1em {
        |q_0\ra & & \gate{H} & \ctrl{1} & \qw       & \qw & \qw \\
        |q_1\ra & & \qw      & \targ    & \ctrl{1}  & \qw & \qw \\
        |q_2\ra & & \gate{X} & \qw      & \targ     & \qw & \qw \\
    }$$

    \item Case $\{x_{110}, x_{001}\}$:
    $$\Qcircuit @C=1em @R=1em {
        |q_0\ra & & \gate{X} & \targ        & \qw       & \qw & \qw \\
        |q_1\ra & & \gate{H} & \ctrl{-1}    & \ctrl{1}  & \qw & \qw \\
        |q_2\ra & & \qw      & \qw          & \targ     & \qw & \qw \\
    }$$
\end{enumerate}

\subsubsection*{Three-component superpositions}

To probe reconstruction performance beyond a single MST edge, we also consider
states with $k=3$ nonzero amplitudes. (See Figure~\ref{fig:support3})
These benchmarks correspond to small trees with two edges, allowing us to
examine how multiple interference steps combine under realistic noise.
The preparation circuits use single-qubit rotations of the form
$U(\pi/4,0,0)$ to generate unequal amplitude weights, followed by controlled
operations that entangle the rotated qubits and distribute amplitude across
three computational-basis states.
The resulting supports include both low- and higher-Hamming-distance edges,
producing MSTs with heterogeneous edge weights.

Representative circuits include:
\begin{enumerate}
    \item Case $\{x_{000}, x_{010}, x_{100}\}$:
    $$\Qcircuit @C=1em @R=1em {
        |q_0\ra & & \qw                           & \qw       & \qw      & \qw & \qw \\
        |q_1\ra & & \gate{U(\frac{\pi}{4}, 0, 0)} & \targ     & \qw      & \qw & \qw \\
        |q_2\ra & & \gate{U(\frac{\pi}{4}, 0, 0)} & \ctrl{-1} & \gate{H} & \qw & \qw
    }$$

    \item Case $\{x_{000}, x_{010}, x_{101}\}$:
    $$\Qcircuit @C=1em @R=1em {
        |q_0\ra & & \qw                           & \qw       & \qw      & \targ      & \qw \\
        |q_1\ra & & \gate{U(\frac{\pi}{4}, 0, 0)} & \targ     & \qw      & \qw        & \qw \\
        |q_2\ra & & \gate{U(\frac{\pi}{4}, 0, 0)} & \ctrl{-1} & \gate{H} & \ctrl{-2}  & \qw
    }$$

    \item Case $\{x_{000}, x_{011}, x_{110}\}$:
    $$\Qcircuit @C=1em @R=1em {
        |q_0\ra & & \qw                           & \qw       & \targ     & \qw       & \qw \\
        |q_1\ra & & \gate{U(\frac{\pi}{4}, 0, 0)} & \targ     & \ctrl{-1} & \targ     & \qw \\
        |q_2\ra & & \gate{U(\frac{\pi}{4}, 0, 0)} & \ctrl{-1} & \gate{H}  & \ctrl{-1} & \qw 
    }$$
\end{enumerate}

\subsubsection*{Scaling experiments}

Finally, to study the dependence on system size while keeping the support
geometry fixed, we consider GHZ-type two-component states of the form
\(
\ket{\psi} = \tfrac{1}{\sqrt{2}}(\ket{0^{\otimes n}} + \ket{1^{\otimes n}}),
\)
with $n$ ranging from four to ten qubits.
These states have support size $k=2$ but maximal Hamming distance
$h = n$, corresponding to a single MST edge whose weight grows linearly with
system size.
Each state is prepared using a linear cascade of CNOT gates originating from a
single Hadamard gate, ensuring minimal circuit depth and a fixed entanglement
pattern across all system sizes.

\begin{enumerate}
    \item Case $\{x_{0000}, x_{1111}\}$:
    $$\Qcircuit @C=1em @R=1em {
        |q_0\ra & & \gate{H}    & \ctrl{1}  & \qw       & \qw       & \qw \\
        |q_1\ra & & \qw         & \targ     & \ctrl{1}  & \qw       & \qw \\
        |q_2\ra & & \qw         & \qw       & \targ     & \ctrl{1}  & \qw \\
        |q_3\ra & & \qw         & \qw       & \qw       & \targ     & \qw
    }$$

    \item Case $\{x_{00000}, x_{11111}\}$ and 
          $\{x_{000000}, x_{111111}\}$ are prepared analogously via extended
          CNOT cascades:
    (four- and five-qubit circuits shown here)
    $$\Qcircuit @C=1em @R=1em {
        |q_0\ra & & \gate{H}    & \ctrl{1}  & \qw       & \qw       & \qw       & \qw \\
        |q_1\ra & & \qw         & \targ     & \ctrl{1}  & \qw       & \qw       & \qw \\
        |q_2\ra & & \qw         & \qw       & \targ     & \ctrl{1}  & \qw       & \qw \\
        |q_3\ra & & \qw         & \qw       & \qw       & \targ     & \ctrl{1}  & \qw \\
        |q_4\ra & & \qw         & \qw       & \qw       & \qw       & \targ     & \qw
    }$$
\end{enumerate}

\end{document}